# Economic geography and the scaling of urban and regional income in India

Anand Sahasranaman[1,2,*] and Luís M. A. Bettencourt[3,4,#]

[1]Division of Mathematics and Computer Science, Krea University, Sri City, AP 517646, India.

[2]Centre for Complexity Science, Dept of Mathematics, Imperial College London, London SW72AZ, UK.

[3]Mansueto Institute for Urban Innovation, Dept of Ecology and Evolution, Dept of Sociology, University of Chicago, Chicago IL 60637, USA.

[4]Santa Fe Institute, Santa Fe NM 87501, USA

[*] Corresponding Author. Email: anand.sahasranaman@krea.edu.in

[#] Email: bettencourt@uchicago.edu

**Abstract:**

We undertake an exploration of the economic income (Gross Domestic Product, GDP) of Indian districts and cities based on scaling analyses of the dependence of these quantities on associated population size. Scaling analysis provides a straightforward method for the identification of network effects in socioeconomic organization, which are the tell-tale of cities and urbanization. For districts, a sub-state regional administrative division in India, we find almost linear scaling of GDP with population, a result quite different from urban functional units in other national contexts. Using deviations from scaling, we explore the behavior of these regional units to find strong distinct geographic patterns of economic behavior. We characterize these patterns in detail and connect them to the literature on regional economic development for a diverse subcontinental nation such as India. Given the paucity of economic data for Urban Agglomerations in India, we use a set of assumptions to create a new dataset of GDP based on districts, for large cities. This reveals superlinear scaling of income with city size, as expected from theory, while displaying similar underlying patterns of economic geography observed for district economic performance. This analysis of the economic performance of Indian cities is severely limited by the absence of higher-fidelity, direct city level economic data. We discuss the need for standardized and consistent estimates of the size and change in urban economies in India, and point to a number of proxies that can be explored to develop such indicators.

## Introduction:

The economic performance of India is critical to the well-being of over one-sixth of the world's population. As India continues to urbanize, and over half the population becomes urban over the next few decades [1], cities will become ever more central to India's dynamics of economic growth and human development. Therefore, the need to develop a scientific understanding of the economic performance of Indian regions and cities becomes critical. Our previous work explored the quantitative characteristics of crime, innovation, spatial density and a number of services in Indian cities using the framework of urban scaling [2]. Here, we extend this analysis to regional patterns of economic performance using district and state Gross Domestic Product (GDP) information. Within the constraints imposed by existing units of analysis in the data, we also attempt a systematic exploration of urban GDP, which has been an issue of long-standing interest for Indian cities.

There is a vast literature in economics on the determinants of regional performance and mechanisms of economic growth. Some of the significant drivers identified in this literature include transportation and market access [3–5], agglomeration economies [6–10], and issues related to human capital and its mobility [11–16]. New Economic Geography posits that access to markets, in the form of transportation infrastructure networks, is critical to the trajectory of productivity and wages in sub-national regions [3]. This finding has been found empirically sound across national contexts, including in developing economies such as China and India [4,5]. Specifically, economic potential in India is found to be strongly clustered by geography, with the states of Tamil Nadu, Kerala, and Haryana having the highest concentration of districts with high economic potential, and the state of Uttar Pradesh containing districts with significant economic underperformance [5]. Agglomeration economies in sub-national regions are a measure of preferential attachment effects that are reflected in increasing economic densities and urbanization [6,7]. For instance, the concentration of firms in similar industries is both a cause and a consequence of geographically proximate investments in businesses, creation of local talent pools with string matching, and the realization of knowledge spillovers – and evidence of such agglomeration effects has been empirically validated across nations [8,9]. There is evidence for agglomeration effects in India emanating from inter-industry urbanization economies at the regional level [10]. Human capital is also found to influence levels of productivity through multiple channels [11,12] – with robust evidence available for transmission channels such as the ability created by locally available trained and skilled workforces, knowledge spillovers enabling maximal exploitation of agglomeration economies, and also the possibility of high quality human capital being able to adjust to longer-term structural changes in the economy [13–15]. Empirical work suggests that human capital (education) has been a significant contributor to increase in output per worker in India [16].

There is also a significant body of literature on the linkages between urbanization and the economy. We find that urban locations enable concentration of economic activity through access to diverse labor pools which enable specialization [17,18], reduced costs on account of proximity to users and suppliers as well as cheaper transport [3], speedy and effective responses to changing market conditions [19,20], and enhanced potential for innovation due to geographically concentrated availability of educated and creative human capital [21]. It has also been empirically shown, based on cross-country data, that the rate of urbanization at national level exhibits strong positive correlation with GDP per capita [22]. This strong positive correlation between per capita income levels and urbanization has been observed to be robust even at a more granular, sub-national level based on state-level data in India [23]. However, it is important to recognize that while there is a clear and robust relationship between urbanization and income levels, this does not necessarily translate into a causal relationship. Indeed, multiple empirical studies on this question find no *systematic* relationship between urbanization and economic growth [22,24,25]. Therefore, it appears that while urbanization is part of the economic development process, there is no evidence that it, per se, independently and causally impacts economic growth.

Given this context, our attempt in this work is to explore economic growth in Indian regions and cities using the framework of urban scaling and urban geography [26–28]. Scaling uses population size as the basis for isolating general agglomeration effects, specifically characterized by increasing returns to scale in socioeconomic interactions such as innovation and GDP in cities – as evinced in empirical studies across multiple national jurisdictions [2,27,29–31].

In scaling analysis, an indicator such as GDP, $Y_i(t, N_i)$, for city $i$, with population size $N_i(t)$, at time $t$ is given by:

$$Y_i(t, N_i) = Y_0(t) N_i^\beta e^{\xi_i(t)}, \qquad (1)$$

where $Y_0(t)$ is a measure of systemic change in GDP across all regions in the analysis, independent of population size with dimensions of a flow of money per year (income).

The scaling exponent $\beta$ is the elasticity of $Y_i$ to population size. This parameter, when measured for urban functional areas, is found to fall in three distinct universality classes containing attributes that represent socioeconomic interactions ($\beta > 1$), economies of scale ($\beta < 1$), and individual human needs ($\beta \simeq 1$) respectively [27]. Empirically, across a range of countries, it has been observed that $\beta \simeq 7/6 > 1$ for the GDP indicator [27,29,31], which is also in line with theoretical expectation [26]. Eq. (1) also allows us to express the average dynamics of a set of units, via the temporal change in the *centres* for the data in logarithmic variables ($\langle \ln Y(t) \rangle, \langle \ln N(t) \rangle$), defined by the average of a set of units as

$$\langle \ln Y(t) \rangle = \frac{1}{N_c} \sum_{i=1}^{N_c} \ln Y_i(t), \quad \langle \ln N(t) \rangle = \frac{1}{N_c} \sum_{i=1}^{N_c} \ln N_i(t), \qquad (2)$$

where $N_c$ is the total number of units (cities or regions) in the set.

The quantities $\xi_i(t)$ are specific to individual cities or regions ($i$) and represent the local, idiosyncratic features that affect their GDP away from the scaling average. Specifically, $\xi_i(t)$ represent scale (population-size)-independent deviations of individual regions from the scaling relation:

$$\xi_i(t) = \ln \frac{Y_i}{Y_0 N_i^\beta}. \qquad (3)$$

Using this simple but systematic framework of scaling analysis and its deviations, we explore regional economic growth in India, measured as district level GDP (Gross District Domestic Product or GDDP) and examine the resulting economic geography in light of empirical evidence from regional economic analyses. We also attempt to use this type of (non-explicitly urban) regional data to approximate the set of all largest Indian cities and examine the nature of resulting scaling relationships and urban economic geography.

## 2. Scaling and Economic Geography of District GDP in India

We use publicly available data for GDP income for districts across 12 Indian states, accounting for over 74% of the country's population. For a detailed description of data sources and statistical methods, please refer Appendix A.

We start by exploring the simplest scaling relationship, between district GDP (GDDP) and corresponding district population size. Districts tile the entire territory of India and thus vary enormously in character, some may be parts of large cities as we shall see below, while others will encompass together rural areas and towns. Nevertheless, we can analyze their scaling relation with population, essentially asking if these units of analysis somehow manifest any increasing returns in GDP per capita with their population as "agglomeration effects".

We find that GDDP scaling is at the cusp of linearity and superlinearity (Figure 1A), with $\beta = 1.02$ (95% Confidence Interval (CI): 0.92, 1.13). This relationship is very noisy and quite distinct from expectations for functional urban areas (defined as integrated labor markets), for which $\beta$ is expected to be in the region of 7/6 as observed in other nations [27,29,31]. Taking this result at face value tells us that Indian districts are not, on average over all district types and regions, generating agglomeration economies (as measured by GDDP), as predicted by urban scaling theory [26]. We also plot the rank order of deviations from scaling law, $\xi_i(t)$ (Figure 1B), which are dimensionless and enable direct comparison between districts due to the exclusion of population size effects.

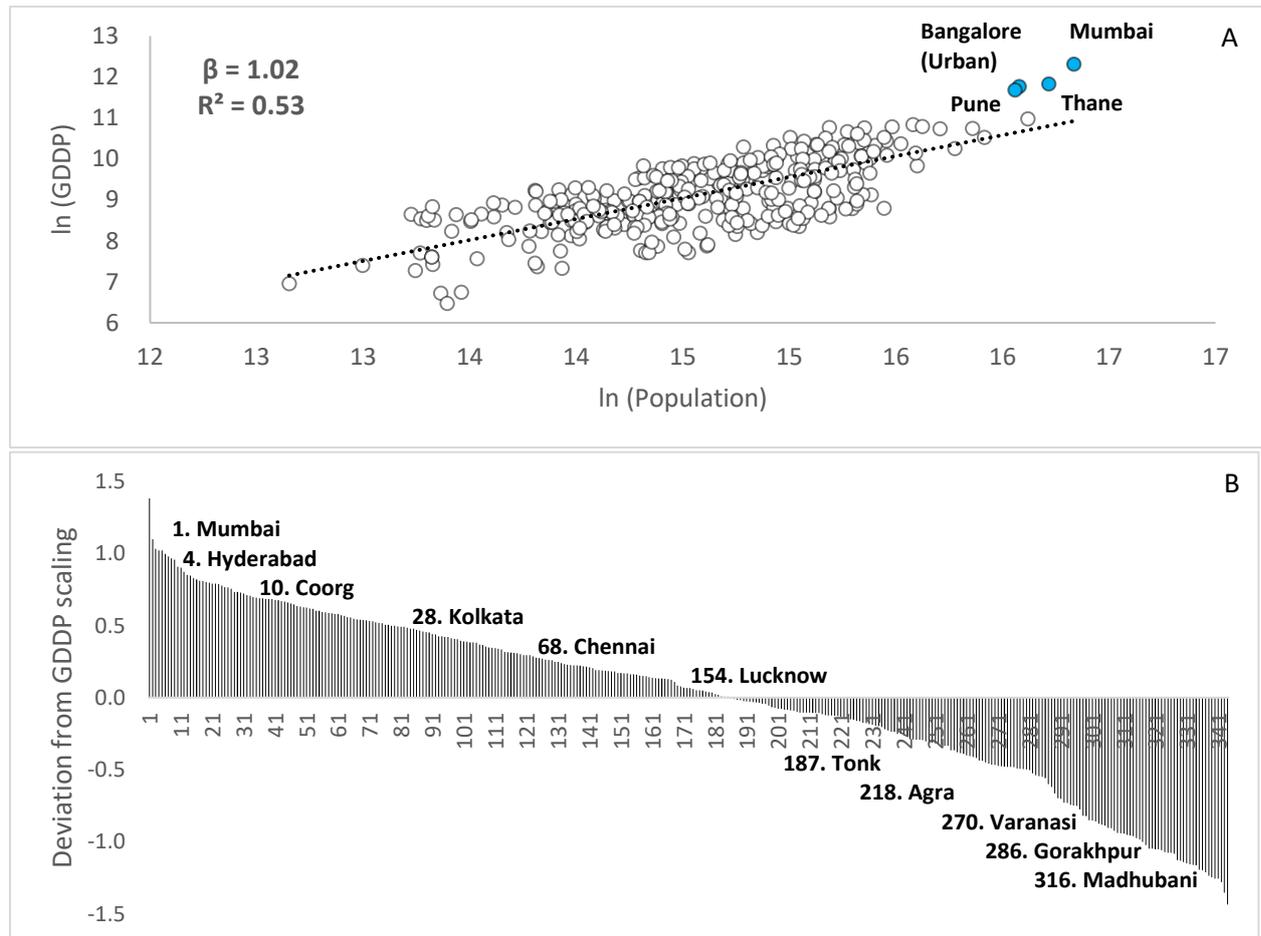

*Figure 1*: **Scaling and deviations of GDDP of Indian districts (2011)**: A: Scaling of district level Gross Domestic Product (GDDP) with district population (2011). This shows a very slight - not statistically significant - superlinear relationship to population with an exponent of 1.02 (95% Confidence Interval (CI): [0.92, 1.13]). B: Rank order of districts based on deviations from scaling relation, $\xi_i(t)$, depicting the local, idiosyncratic effects of individual districts in driving scaling behaviour. We see that districts corresponding to large Indian cities (Mumbai and Thane districts contained within the Mumbai Urban Agglomeration, Bangalore Urban district in the Bangalore Urban Agglomeration, and Pune district in the Pune Urban Agglomeration) are strong positive outliers, with much larger economies per capita than the average scaling trend would predict.

In our previous analysis of the properties of Indian urban agglomerations [2], it emerged that the diverse geography of India was critical to the emergence of scaling behaviour, as far as crime and, to a lesser extent, technological innovation (measured by patenting activity) are concerned. Specifically, we found that cities in north-central and eastern India performed qualitatively and quantitatively in distinct ways from the cities in southern and western India. In order to elicit a scientifically robust understanding of the economic geography of India, we analyse the deviations from the scaling relation for GDDP, $\xi_i(t)$ (from Eq. 2 and Figure 1B), which are scale-independent and provide a principled mechanism to characterize regional effects. When we map $\xi_i(t)$ for district GDDP in the year 2011, we obtain a spatial distribution of deviations as shown in Figure 2, which provides a visual representation of geographic disparities in economic performance across the country. It clearly emerges that districts in north-central and eastern India, on average, tend to underperform the average scaling expectation (dashed line in Figure 1A) as highlighted by the clustering of red circles in these regions, while districts in southern and western India, on average, display overperformance as evinced by the predomination of blue circles.

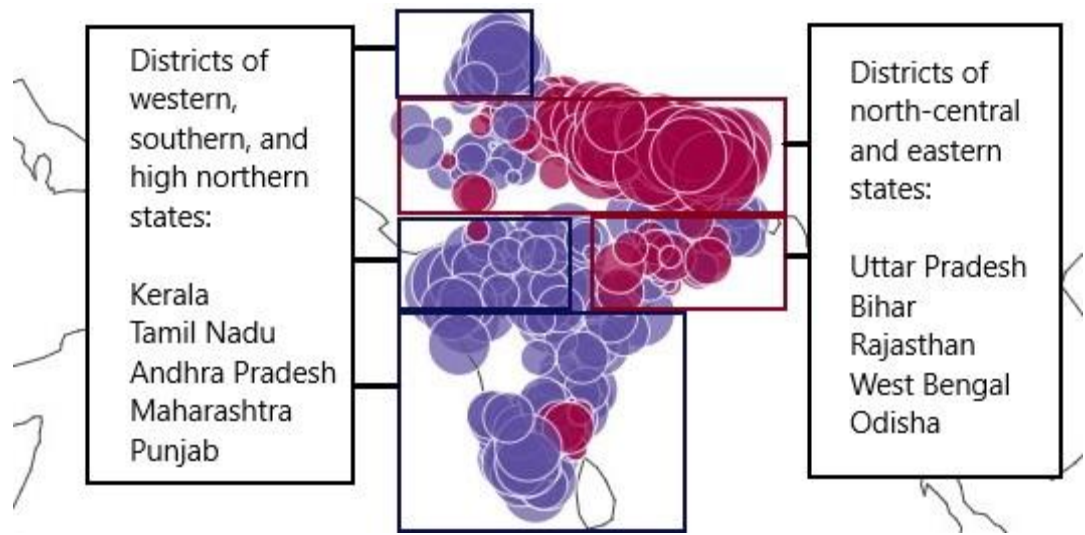

*Figure 2*: **Economic geography according to deviations from District GDDP scaling law ($\xi_i(t)$, 2011)**: Each circle represents the deviation, $\xi_i$, of a single district. Blue circles represent over-performance with respect to scaling law and red circles under-performance. The size of the circle indicates the magnitude of over- or under-performance. Blue circles are concentrated in the south, west, and high north of the country, while red circles are predominant in north-central and eastern part (Indo-Gangetic plain) of the country.

While Figure 2 strongly suggests an underlying pattern of regional economic geography in India, we seek to validate this impression more formally by clustering districts based on the distance between the time-series of their deviations from scaling. Figure 3A shows a heatmap of the Euclidean distance between pairs of time-series of deviations: The closer to zero this distance is, the greater the similarity in the temporal evolution of deviations. This heatmap suggests a total of seven clusters (Figure 3A). Clusters 1 and 2 (Figure 3A&B) are composed of severely underperforming (with respect to scaling law) districts from the north-central states of Uttar Pradesh and Bihar, respectively. Clusters 3 and 4 are predominantly composed of overperforming districts in the southern, western, and high northern parts of the country. Cluster 6 is composed of underperforming districts from north-central and eastern India, while Cluster 7 consists of districts from the same region that on average perform in accordance with scaling expectations. Finally, Cluster 5 is a remainder, a mixed cluster composed of districts from across India, which on average slightly overperform the expectations from the scaling law. Overall, the composition of these clusters formalizes the notion of regional economic geography hinted at in Figure 2, with districts in north-central and eastern India (representing the states of Uttar Pradesh, Bihar, Rajasthan, Odisha, West Bengal in our data set) lagging on economic performance, while the districts in the south, west, and high north (comprising the states Tamil Nadu, Kerala, Andhra Pradesh, Maharashtra, and Punjab) consistently exceed scaling expectations. This result also finds close agreement with empirical evidence that the north-central and eastern states have historically (since 1947) showed lower comparative socioeconomic development as represented by evolution of GDP and other social indicators (literacy, mortality, population growth), when compared to the significantly better performance of southern and western states [32–35]. More recently, assessing the performance of Indian districts on the Multidimensional Poverty Index (MPI) [36], which measures serious deficits in health, education, and living standards, it emerges that 91 out of the 100 districts with worst MPI in the country are in the north-central and eastern states (Bihar, Uttar Pradesh, Odisha, Rajasthan, Madhya Pradesh, Jharkhand, and Chattisgarh).

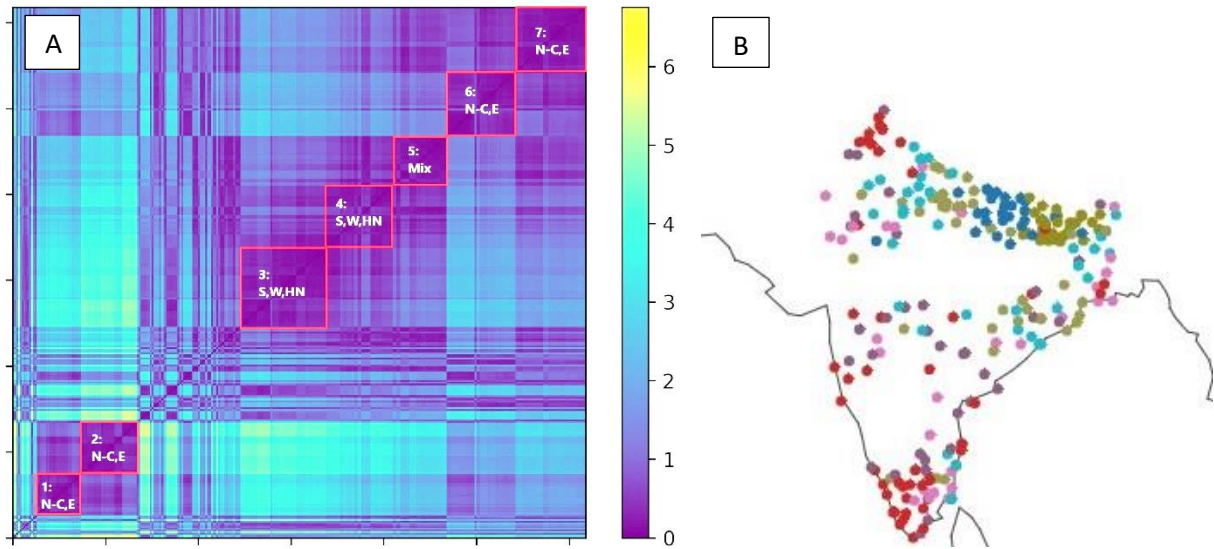

*Figure 3*: **Clustering Analysis:** A: Heatmap of time-series of deviations from District GDDP scaling relation ($\xi_i(t)$, 2006 to 2011). Clusters are based on Euclidean distance between time-series (from 2006 to 2011) of pairs of district GDDP deviations. A total of seven clusters are clearly separable, which show a clear regional economic geography, with underperforming districts primarily belonging to the north-central and eastern parts of India, while the overperformers predominant in the south and west. The clusters are identified by the regions their districts primarily belong to: N-C: North-Central (representing the states of Uttar Pradesh, Bihar, Rajasthan), E: East (West Bengal, Odisha), S: South (Tamil Nadu, Kerala, Andhra Pradesh), W: West (Maharashtra), HN: High North (Punjab), and Mix: Mixture of districts from across the nation. B: Map of Clusters. The map depicts the geographical spread of the seven clusters that emerge out of the clustering analysis. Color Code: Dark Blue: Cluster 1, Dark Green: Cluster 2, Red: Cluster 3, Purple: Cluster 4, Pink: Cluster 5, Light Green: Cluster 6, Light Blue: Cluster 7.

Despite the confirmation of this geographical pattern to economic performance, the clustering analysis also points to differentiated performance within geographies. For instance, districts in the states of Bihar and Uttar Pradesh (in north-central India) ought to be of particular concern to policy makers because almost all districts in these states (97% in Bihar or 37 out of 38 districts, and 89% in Uttar Pradesh or 62 out of 70 districts) have significant, negative deviations from scaling (with state-level average GDDP deviations of -0.98 and -0.43 respectively), concentrated in Clusters 1 and 2 in Figure 3. Other states in the north-central and eastern region such as West Bengal and Rajasthan, while still having a significant proportion of districts underperforming (47% in West Bengal or 9 out of 19 districts, and 38% in Rajasthan or 12 out of 32), however have a majority of their districts overperforming the scaling relation, consequently yielding state level average GDDP deviations in the region of ~0.10, and find themselves clustered in Clusters 5, 6 and 7 (Figure 3). In the better performing southern, western, and high northern regions of the country, we find that only 6 out of the 120 districts (5 in Tamil Nadu and 1 in Maharashtra) underperform the scaling law. Some of these intra-geographic differences become apparent when we plot the temporal evolution of the centres of the population-GDDP distributions (log-log scale) for each state for each year data is available (Eq. 2).

As Figure 4A clearly illustrates the GDDP centres of the Bihar and Uttar Pradesh distributions are the lowest amongst all states (even as they show an increasing trend), with Odisha's GDDP centre a little higher than these two states, and Rajasthan and West Bengal showing the highest GDDP centres in the north-central and eastern region. These intra-geographic differentiations also echo some of the economic geography findings of Bhandari and Khare [37], whose economic model of district performance finds that districts in Uttar Pradesh and Bihar show a significant decline in their share of the economy over time, while districts in Rajasthan show increase in economic share. It is also however apparent from Figure 4A that the GDDP centres of all the southern, western and high northern states are significantly higher than those of even Rajasthan and West Bengal (even Karnataka whose temporal evolution of GDDP centre appears very similar to Rajasthan, is doing so at lower population centre)

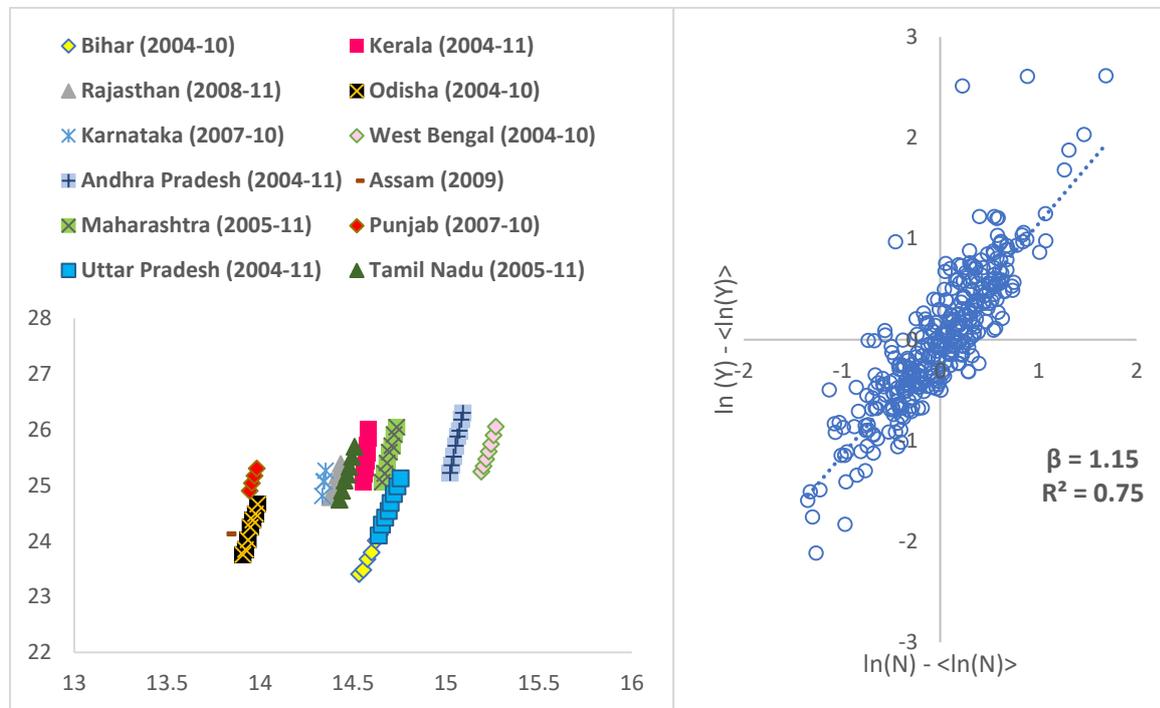

*Figure 4*: **Centering Analysis:** A: Temporal evolution of Population-GDDP centres per state per year: The centre for each state for each year is calculated as Eq. (2) for districts in each state in each given year. The north-central states of Uttar Pradesh and Bihar have the lowest GDDP centres despite having relatively high population centres. The southern and western states show the highest GDDP centres over time. B: Centred Scaling of GDDP (2010): The scaling of GDDP with population when the data have been centred in each state shows superlinear scaling of GDDP, with an exponent of 1.15 (95% Confidence Interval (CI): [1.08, 1.22]).

India's federal structure has ensured that states have significant powers in the design and implementation of social and economic policy [34], and the heterogenous economic paths charted by different states post 1947 are testament to the decision making powers of Indian states. Given this underlying reality where the baseline GDP of different states shows significant variation, the centred scaling relationship (Eq. 2) provides us with a single-parameter model to estimate the scaling exponent, while excluding baseline state differences. Figure 4B is a plot of the centred scaling of GDDP with population, and this reveals an exponent of 1.15 (95% CI of [1.08,1.22]), which is in reasonable agreement with the expected exponent of 7/6 from urban scaling theory [27]. This starts to suggest the presence of agglomeration effects at the district level, which can be masked by regional disparities

We explore this phenomenon further by splitting the GDDP data set into two sets, based on the geography suggested by this analysis of deviations. Figure 5 shows the quantitatively distinct scaling relationships exhibited by these two sets of districts segregated by geography.

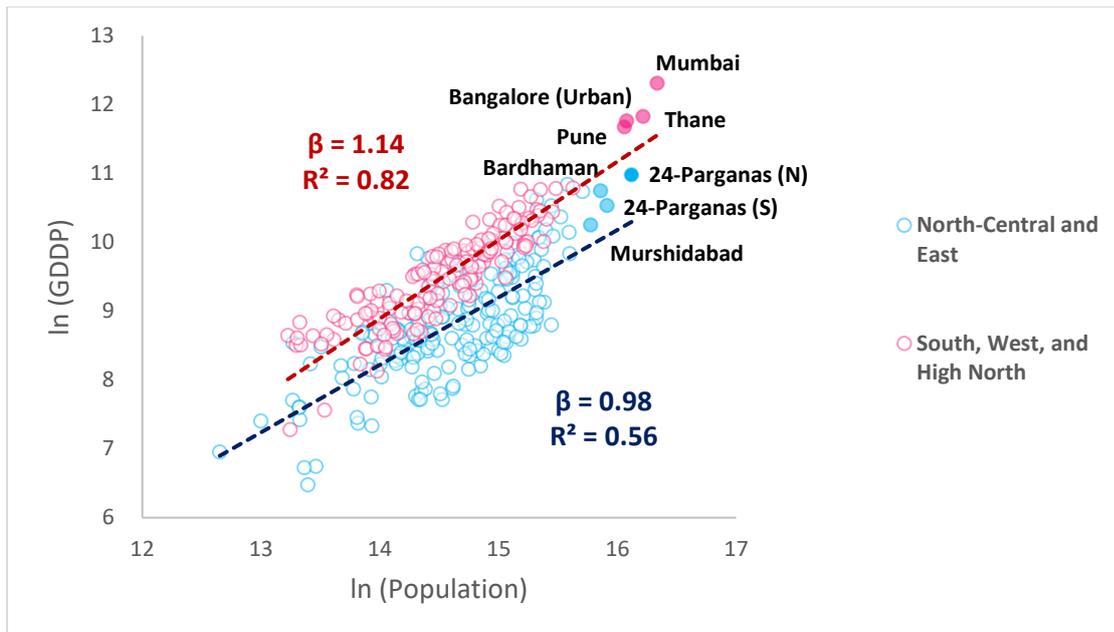

*Figure 5*: **Scaling of Indian districts by geography (2011)**: Scaling of district Gross Domestic Product (GDDP) with population for two sets of data. The pink data points (Set 1) represent the scaling of districts in the south, west, and high north of India (comprising the states Karnataka, Tamil Nadu, Kerala, Andhra Pradesh, Maharashtra, Punjab) and the blue data points (Set 2) are districts in north-central and eastern India (comprising the states Uttar Pradesh, Bihar, Odisha, Rajasthan, West Bengal). There is a significant difference in the nature of scaling relationships, with Set 1 exhibiting a superlinear relationship in line with expectation from scaling theory (exponent of 1.14 with a 95% CI of [1.05, 1.22]), while Set 2 shows a slightly sublinear relationship (exponent of 0.98 with a 95% CI of [0.86, 1.10]). Again, we see in both Set 1 and Set 2 that the largest districts outperform the respective scaling lines - Mumbai, Thane, Bangalore Urban and Pune in Set 1, and 24-Parganas (N), 24-Parganas (S) (both of which are part of the Kolkata Urban Agglomeration), Bardhaman, and Murshidabad (all of these districts belong to the state of West Bengal) in Set 2.

We observe, again, that districts in the southern, western, and high northern states (Karnataka, Tamil Nadu, Kerala, Andhra Pradesh, Maharashtra, Punjab) show a superlinear scaling relationship of GDDP with $\beta = 1.14$ (95% CI: 1.05, 1.22), which is in keeping with expectations from empirical observations elsewhere [27,29,31] as well as theory [26]. On the other hand, economic performance in districts of north-central and eastern states (Uttar Pradesh, Bihar, Rajasthan, Odisha, West Bengal) shows slightly sublinear scaling, $\beta = 0.98$ (95% CI: 0.86, 1.10), suggesting the absence of advantages of scale deriving from socioeconomic interactions in these regions.

## 3. Scaling and Urban Geography of Income in India

We now attempt to understand scaling of GDP in the context of Indian cities.

We begin by exploring the relationship between *state level* GDP (Gross State Domestic Product, or GSDP) and its urban population. We find a sublinear scaling relationship, with an exponent $\beta = 0.90$ (Figure 6). While we might expect a superlinear relationship with increasing urban population, the observed result is potentially explained by the effect of the highly populous, low-income states of north and central India, which despite low levels of urbanization have high overall urban population counts. For instance, the states of Bihar, Uttar Pradesh, and Rajasthan rank at 30, 26, and 22 respectively (out of 31 states) in terms of their urbanization levels, but in terms of urban population counts, they rank 12, 2, and 9 respectively.

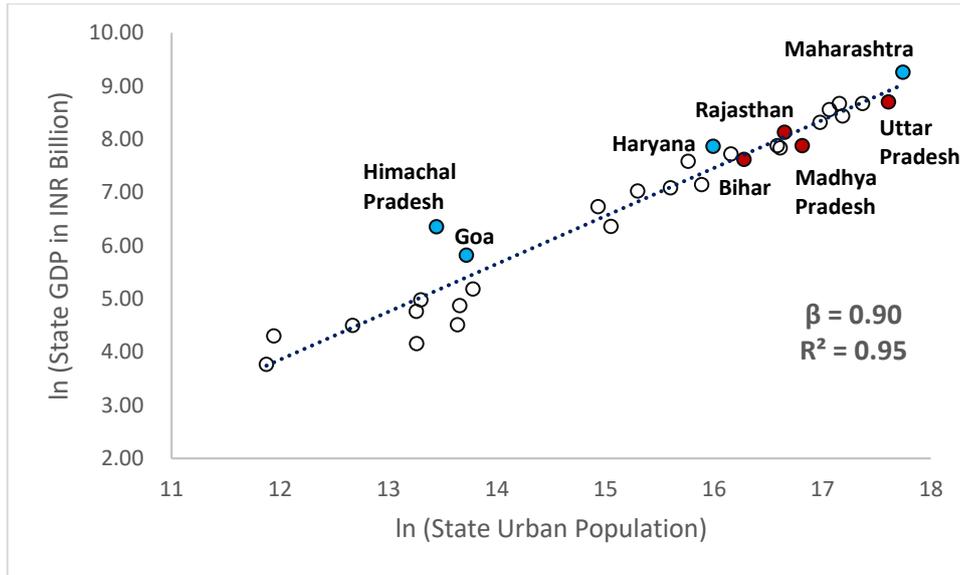

*Figure 6*: **Scaling of GSDP with state urban population (2011)**: A: Scaling of state level Gross Domestic Product (GSDP) with state urban population (2011). This shows a sublinear relationship with an exponent of 0.90 (95% Confidence Interval (CI): [0.82, 0.98]). The large states of Bihar, Uttar Pradesh, and Madhya Pradesh have low levels of urbanization, but comparatively high urban populations and underperform relative to the scaling line. Smaller states like Haryana, Himachal Pradesh, and Goa overperform relative to the scaling line.

We have however seen that there exists a robust, empirically tested relationship between urbanization and per capita incomes, both at a cross-country level and at a cross-state level within a country [22,23]. We now drill down from the state to the district level and find that this positive relationship between per capita income and urbanization obtains even at this level of granularity, based on Indian data (Figure 7A). We also seek to understand how this relationship compares with the relationship between the deviations from GDDP scaling law, $\xi_i(t)$, and urbanization. Given that per capita income increases with urbanization, we would expect that $\xi_i(t)$ would capture the effect of increasing income in more urbanized districts. Indeed, as Figure 7B illustrates, we find that there is a positive relationship between $\xi_i(t)$ and urbanization. Overall, the two curves indicate not only a close qualitative concurrence but also a quantitative one in terms of the functional forms of the best fit curves that describe the relationships of $\xi_i(t)$ and per capita income with urbanization.

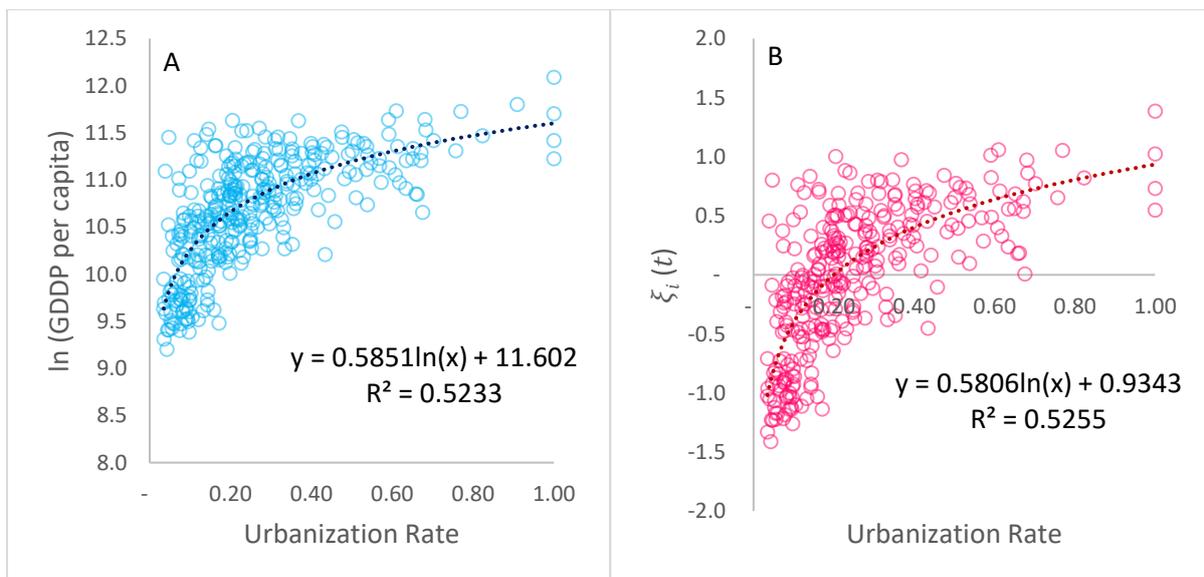

*Figure 7*: **Income and Urbanization**: A: ln (GDDP per capita) v. District urbanization rate. There is a positive relationship between urbanization and income per capita at the district level. B: Deviation from GDDP scaling law ( $\xi_i(t)$ in Figure 1) v. District urbanization rate. $\xi_i(t)$ also displays a similarly positive relationship with urbanization. The functional forms of the curves of best fit in both cases show very close correspondence, with almost the same coefficient (0.5851 and 0.5806); the normalization is clearly different, however.

This naturally leads to the ultimate question of how income scales across urban agglomerations. As we highlighted in earlier work [2], there is a lack of systematic collection and dissemination of economic data at the level of Indian cities and therefore, in terms of official statistics, we are left with using district level GDP as a proxy for city GDPs. In order to create a dataset of such proxied city GDPs we start with considering districts that are predominantly urban, i.e. with urbanization rates of at least 50%. Given that per capita urban incomes in India are, on average, 2.75 times per capita rural incomes [38], it is a reasonable assumption that a very significant proportion of the GDP in majority urban districts is produced by the corresponding urban components. This leaves us with a set of 38 districts, out of which we only consider those districts in which there is a single identifiable city that contributes significantly (over 65%) to the urbanization of that district. On average, we find that the final set of proxied cities thus obtained, contribute to over 86% of the urbanization of their districts. We also have four urban agglomerations in the data – Mumbai, Kolkata, Chennai, and Hyderabad – that extend across multiple districts and in these cases, we aggregate the population and GDP of the constituent districts to proxy the data for these cities, see Appendix A. We also obtained data for Delhi state, which closely corresponds to the Delhi Urban Agglomeration, and incorporate this as an urban unit into the analysis. Overall, the final dataset thus created has 24 urban areas, which we use for analysis of scaling and deviations, see Figure 8. It is apparent that the creation of even this limited dataset involves several approximations and assumptions (discussed in Appendix A), and while the data clearly do not capture exact representations of functional Indian cities, what it offers us is a starting point (in the absence of better data) to begin to explore urban GDP scaling.

When we plot the scaling of these city GDPs with population (Figure 8A), we find superlinear scaling with an exponent of 1.12 (95% CI: 0.94, 1.30), which is consistent with expectations from functional cities in other nations and from theory [26,27,29,31]. When we compare the rank orders of cities dataset based on per capita GDP and deviations ($\xi_i(t)$) from the scaling law at work here, we find that the largest cities – Mumbai, Delhi, Kolkata, Chennai, Hyderabad, Pune - rank slightly worse (or at best, the same) under $\xi_i(t)$ than per capita GDP, which is explained by the expectations of superlinear increase in GDP with population under scaling (Figure 8B).

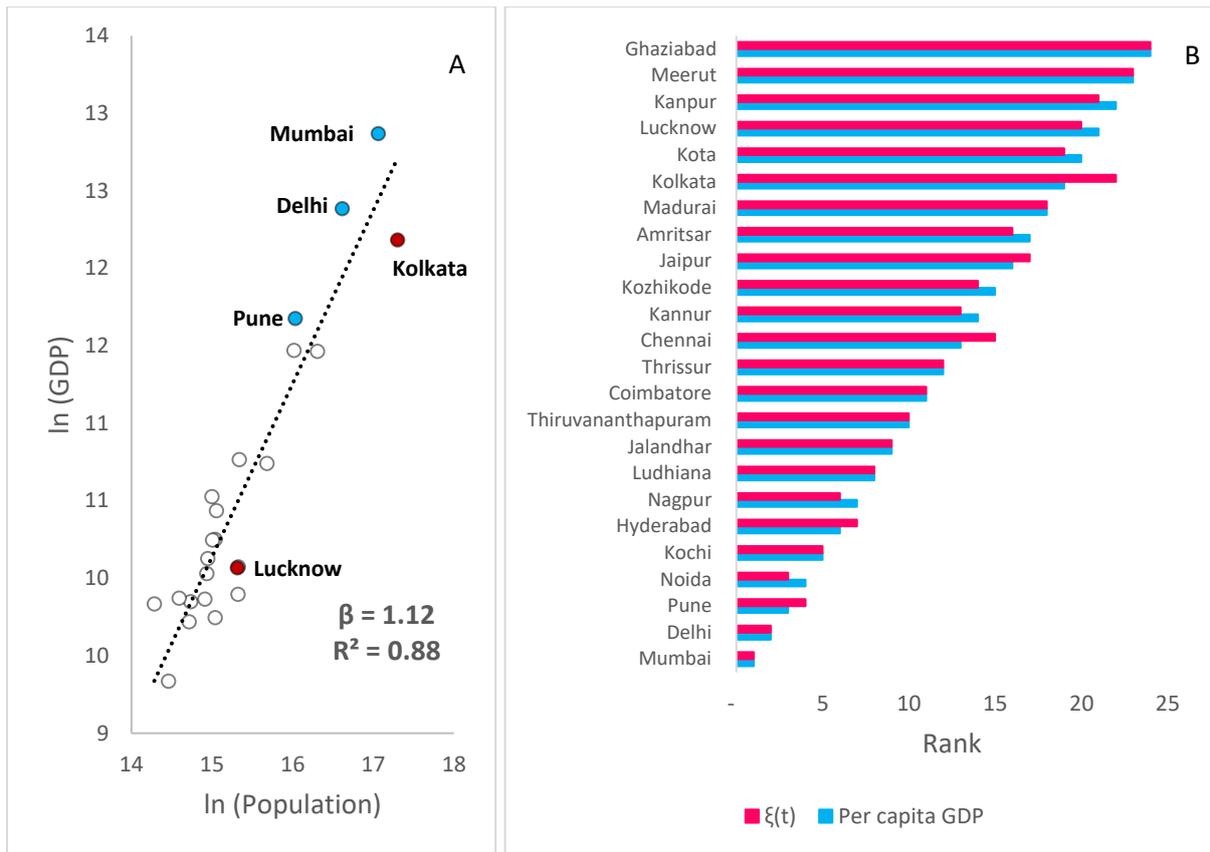

*Figure 8*: **Scaling of Indian cities and Rank order of per capita GDP/scaling deviations (2011)**: A: Scaling of GDP with population reveals a superlinear relationship with exponent β = 1.12 (95% CI: [0.94, 1.30]), which is in line with expectations. Mumbai, Delhi, and Pune outperform the scaling line, while Kolkata and Lucknow underperform. B: Ranking large Indian cities by per capita GDP and $\xi_i(t)$, we find that $\xi_i(t)$ rankings are slightly poorer for the largest cities under $\xi_i(t)$ than per capita GDP.

We now turn to exploring the economic geography of urban income using the deviations from scaling, $\xi_i(t)$, as the basis for urban areas rather than districts. The simple visualization in Figure 9A suggests a similar geographical breakup in terms of GDP performance as we saw in the case of district GDDP. On average, cities in the south, west, and high north appear to outperform the scaling law, while those in north-central and east India underperform.

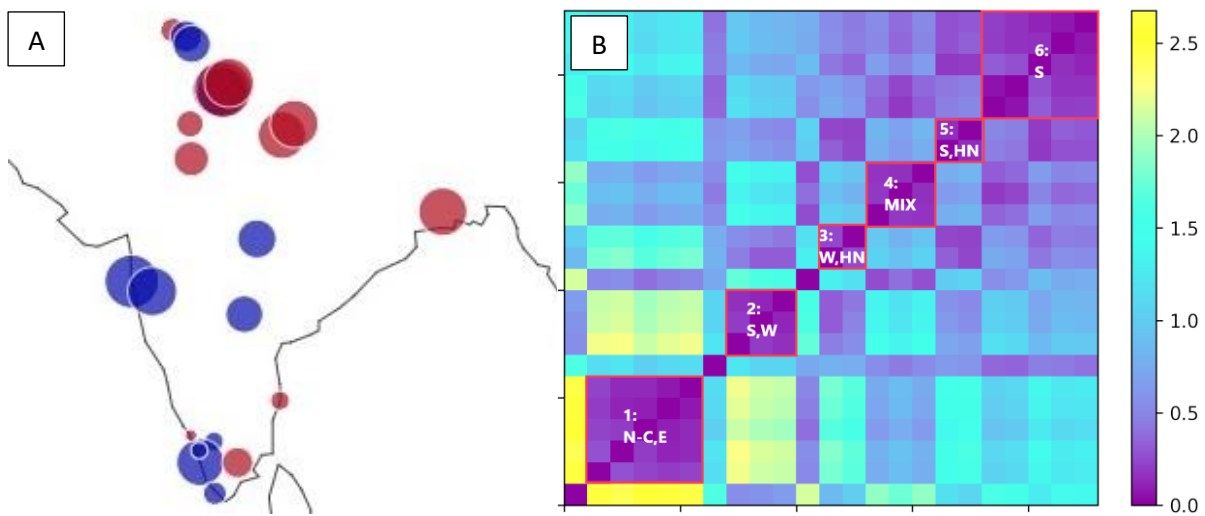

*Figure 9*: **Economic geography and Heatmap based on $\xi_i(t)$**: A: Each circle represents the deviation of a single city from the scaling law. As before, blue circles represent over-performance and red circles under-performance. The size of the circle indicates the magnitude of over- or under-performance. Blue circles are concentrated in the south, west, and high north of the country, while red circles are predominant in north-central and eastern part (Indo-Gangetic plain) of the country. B: Heatmap of time-series deviations from scaling law. Clusters are based on Euclidean distance between time-series (from 2006 to 2011) of pairs of city GDP deviations. This analysis reveals the underlying geographical pattern of urban GDP through the emergence of geographically distinct clusters of cities in north-central and east India (N-C, E), and of cities in the west, south and high north (W, S, HN).

We again formalize this notion by performing a clustering analysis of the Euclidean distance between time-series of $\xi_i(t)$ for all 24 urban areas in the dataset (Figure 9B) and find that this confirms the urban economic geography suggested in Figure 9A. Clustering shows a clear geographical basis with 4 of the 6 resultant clusters composed of cities from the south, west, and high north, one cluster composed of the poorest underperformers from north-central and eastern India, and finally one mixed cluster, as usual in this type of analysis. Overall, despite the construction of the dataset being based on urbanization levels, the economic geography revealed here appears to almost exactly mirror that of district GDDP. This also suggests the possibility that with the availability of higher quality city data, we might see qualitatively and quantitatively different scaling relationships (i.e. different intercepts) between cities across these geographies, just as was manifested in the case of district GDDP.

## 4. Conclusion

We explored a systematic analysis of existing official regional and urban economic data for income in India with the objective of furthering a stronger scientific understanding of economic performance and agglomeration effects. We used the framework of scaling theory as our point of departure, from which we attempted to characterize the economic geography of India using existing regional GDP data. Based on these analyses, we also proposed approximations for functional definitions of large Indian cities based on collections of districts. We measured associated increasing returns to their population scale, which emerge to be consistent with the behaviour for urban areas in other urban systems and with theory.

There are clearly many limitations to the existing data sources analysed here that will be important to address in the future, if a firmer analysis of the properties of Indian cities and their development are to be assessed over time. The Census of India defines *Urban Agglomerations* as an approximation to urban functional areas in most other nations. It would be important in the future that these units are characterized in terms of their economic make up and performance, not only in terms of their GDP, in ways that are consistent over time and space. The modernization of data collections across the country, including taxes and employment and property records (beyond existing surveys) should allow the nation to leapfrog existing practices and create a modern system based on native talent that is well suited to measure, assess and plan future economic activity in its fast-growing cities.

One of the central difficulties of measuring economic activity in integrated urban economies is their spatial definition. In the United States and other OECD nations, the solution of this problem relies on the consistent assessment of daily commuting flows and their integration of geographic political and civic units into the same unified labour market, known as Metropolitan areas [39,40]. An important task ahead for Indian cities then is the construction of analogous functional units, especially given the current scenario – analogous to US cities - where main Indian cities are growing primarily along peripheries [41].

Measurements of urban metropolitan economies are also becoming more accessible, not only through the modernization of official data records, but also through new proxies available online and through new technologies, including digital mapping and remote sensing, assessments of construction, transportation flows, real estate markets and employment listings. These emerging sources typically display biases towards formal and high-tech sectors of economic activity, but can be complemented by neighbourhood surveys and data collections at the local level in more informal setting, a tradition with great vitality in India. Creating a system that can make use of these traditional and emerging sources of information towards a deeper understanding of human sustainable development in Indian cities is a challenge that directly impacts over one-sixth of the world's population. With increasing urbanization, cities will play an ever more central role in the future of India's economy. Developing a better scientific understanding of their economic development will be critical to ensuring that we fully leverage this process so that the benefits of growth are distributed more fairly and equitably and contribute to global sustainability outcomes.

**Appendix A: Data sources and methods**

*District GDP Data:* District level GDP data was released by the erstwhile *Planning Commission of India* and is available on Government of India's Open Government Data (OGD) platform (https://data.gov.in/). The 2004-05 current price GDP time series for 11 Indian states is available at https://data.gov.in/catalog/district-wise-gdp-and-growth-rate-current-price2004-05. There is some difference in the lengths of the time series available for districts in different states, as follows: Andhra Pradesh, West Bengal, Bihar, Odisha, Kerala, Uttar Pradesh, and Punjab have the entire series from 2004-05 to 2010-11, Maharashtra from 2005-06 to 2010-11, Assam for 2009-10, Rajasthan from 2004-05 to 2009-10, and Karnataka from 2007-08 to 2010-11. We also obtained Rajasthan districts GDP data for 2010-11 from the state government's publication "Estimates of District Domestic Product of Rajasthan 2011-12" available at https://bit.ly/2TD2R2G. For Tamil Nadu, the district GDP time series from 2004-05 to 2010-11 was produced from documents released by the Department of Economics and Statistics, Tamil Nadu. To our knowledge, data for districts in other states was not available for this period from any public source. It is likely that this data exists in documents (hard copies) published by the Departments of Economics and Statistics of these states and needs to be digitized. However, even in the absence of this data, the current dataset of districts covers over 74% of the national population and offers a reasonable starting point for our analysis of GDP scaling and economic geography.

In discussion with several experts on district and state level GDP estimates, it was emphasized to us that different states may possibly use different methodologies. GDP estimates for districts may result from using a combination of components measured at the district level and some attributed from the state level based on available district-wise indicators. Therefore, some of the regional variations in our analysis may also reflect, in addition to variations in actual performance, states' varying statistical practices.

*Urban Area GDP Data:* For all urban areas that were contained within single districts and contributed to over 65% of the district's population, we use the GDP and population measures of the entire district. There are however 4 urban areas that expand across multiple districts: Mumbai Urban Area comprising the districts of Mumbai, Thane, and Raigad; Hyderabad Urban Area comprising Hyderabad and Rangareddy; Chennai Urban Area comprising Chennai, Thiruvallur, and Kanchipuram; and Kolkata Urban Area comprising Kolkata, Haura, 24-Parganas (North), 24-Parganas (South), and Hugli. In each of these cases the GDP and population estimates for the urban areas were obtained as the sum of GDPs and populations of their constituent districts. It is important to realize that adding the GDP of these districts without consideration for intermediate inputs between them may overestimate the total GDP of the set.

In order to include Delhi in the GDP analysis of urban areas, we obtain GDP data for Delhi state (aggregated across all the districts that comprise Delhi), which we include as an approximation for the

Delhi Urban Agglomeration. This data is available from the state government's publication "Socio-economic profile of Delhi 2014-15", accessed at https://bit.ly/2GtVqXN.

*State GDP Data:* State GDP data for 2010-11 was released by the Planning Commission, available at https://bit.ly/2GeUM0X.

*Population and Urbanization Data:* Data on total population, rural population and urban population for both districts and states is available from the Census of India 2011, at http://www.censusindia.gov.in/2011census/population_enumeration.html (Primary Census Abstract Data Tables - India & States/UTs - District Level)